\documentclass[10pt,a4paper,twocolumn,english,3p]{elsarticle}
\usepackage{mathptmx}

\usepackage[T1]{fontenc}
\usepackage[utf8]{inputenc}
\usepackage{amsmath}
\usepackage{graphicx}
\usepackage{esint}

\makeatletter

\pdfpageheight\paperheight
\pdfpagewidth\paperwidth

\providecommand{\tabularnewline}{\\}

\usepackage{babel}
\usepackage{amssymb}
\usepackage{makecell}

\makeatother

\usepackage{babel}
\begin{document}

\title{Universality in the resurgence of generalized Tracy-Widom distributions}

\author{Zoltan Bajnok$^{a}$, Bercel Boldis$^{a,b}$, Dennis le Plat$^{a}$}

\address{\emph{$^{a}$HUN-REN Wigner Research Centre for Physics, 1121 Budapest,
Konkoly-Thege Miklós út 29-33, Hungary}\\
 \emph{$^{b}$Department of Theoretical Physics, Budapest University
of Technology and Economics, Műegyetem rkp. 3., H-1111 Budapest, Hungary}}
\begin{abstract}
We analyze determinants associated with Bessel kernels and generic
symbol functions, which govern a class of observables across all values
of the 't Hooft coupling in supersymmetric gauge theories. Previous
approaches, based on integro-differential equations, have provided
systematic strong-coupling expansions for the logarithms of these
determinants, expressed through the moments of the symbol. The resulting
asymptotic series, once completed with non-perturbative terms, allowed
the leading resurgence relations to be tested. In this Letter we focus
directly on the determinants themselves, and we uncover a strikingly
simple non-perturbative structure: all trans-series corrections share
a universal form, while the moments transform in a direct and intuitive
way. This simplicity points to an underlying organizing principle
for the non-perturbative dynamics of gauge theory observables. 
\end{abstract}
\maketitle

\section{Introduction}

Tracy--Widom distributions \citep{Tracy:1993xj} appear in many areas
of physics and mathematics, providing surprising connections between
seemingly unrelated systems. In particular, distributions related
to Bessel kernels describe eigenvalues of random matrices in the Laguerre
ensemble in an appropriate scaling limit \citep{Forrester:1993vtx}.
Their generalizations have recently emerged in supersymmetric gauge
theories - especially in the context of the AdS/CFT correspondence
\citep{Aharony:1999ti} - yielding exact expressions valid at all
values of the 't Hooft coupling for a variety of observables, including
Wilson loop expectation values \citep{Pestun_2017}, four-point functions
of infinitely heavy BPS operators \citep{Coronado:2018ypq,Coronado:2018cxj,Kostov:2019stn,Bargheer:2019kxb,Kostov:2019auq,Belitsky:2019fan,Bargheer:2019exp,Belitsky:2020qrm,Belitsky:2020qir},
flux-tube correlators \citep{Beisert:2006ez,Belitsky:2019fan,Basso:2020xts}
and many quantities obtained via localization \citep{Beccaria:2020hgy,Beccaria:2021vuc,Billo2022,Beccaria:2023kbl,Korchemsky2025}.
Remarkably, all these observables can be expressed as determinants
of semi-infinite matrices \citep{Beccaria:2022ypy} of the form (\ref{eq:detdef})
.

The mathematical literature contains a long history of studying strong-coupling
expansions of such determinants (see, e.g., \citep{Szego1915,Szego1952,Kac1954,Akhiezer1966,FisherHartwig1969,Bottcher1995OnsagerFisherHartwig,Basor2012BriefHistorySzego}).
More recently, integro-differential equation methods \citep{Its1990}
have enabled systematic expansions to arbitrary order \citep{Belitsky:2020qrm,Belitsky:2020qir}.
These series are asymptotic and require non-perturbative completions
\citep{Beccaria:2022ypy}. Using the analytic structure of the resolvent,
non-perturbative corrections were computed systematically \citep{Bajnok2024,Bajnok20251,Bajnok20252},
and resurgence relations \citep{Marino:2012zq,Dorigoni:2014hea,Aniceto:2018bis}
between the perturbative series and the first few non-perturbative
terms have been verified \citep{Bajnok2024,Bajnok20251,Bajnok20252}.

In this Letter, we go beyond previous studies and analyze the determinants
themselves, uncovering a remarkably simple structure of the full trans-series.
All non-perturbative corrections share a universal form, with only
the moments undergoing a straightforward and intuitive transformation.
Our results reveal a deeper organizing principle for the non-perturbative
dynamics of gauge theory observables and highlight the role of resurgence
in understanding generalized Tracy--Widom distributions.

\section{Observables}

We study the determinant of a semi-infinite matrix 
\begin{equation}
{\cal D}(g)=\det_{1\leq n,m<\infty}(\delta_{nm}-K_{nm}(g))\label{eq:detdef}
\end{equation}
at strong coupling in $g$. The matrix elements depend on the coupling
through the symbol $\chi(x)$ as 
\begin{equation}
K_{nm}(g)=\int_{0}^{\infty}dx\,\psi_{n}(x)\chi\bigl(\frac{\sqrt{x}}{2g}\bigr)\psi_{m}(x)\label{eq:Knm}
\end{equation}
and the basis is formed using the Bessel functions of the same parity
\begin{equation}
\psi_{n}(x)=(-1)^{n}\sqrt{2n+\ell-1}J_{2n+\ell-1}(\sqrt{x})/\sqrt{x}
\end{equation}
which are ortonormal for the inner product in (\ref{eq:Knm}), i.e.
for $\chi=1$. The determinant depends also on $\ell$, which is a
non-negative integer, but we supress this dependence in the notation.
The symbol $\chi(x)=\Theta(1-x)$ describes the Tracy-Widom distribution
\citep{Tracy:1993xj}, which corresponds to the level spacings in
the Laguerre ensemble near the hard edge. By smoothening the cut-off
function we can obtain observables, which appeared recently in various
supersymmtric gauge theories, see table \ref{Table:symbols} to summarize
some of them \citep{Bajnok20251}. 
\begin{table}
\begin{centering}
\begin{tabular}{|c|c|}
\hline 
Wilson loop  & \rule{0pt}{2.5ex}$-\frac{4\pi^{2}}{x^{2}}$\tabularnewline
\hline 
flux tube correlator  & \rule{0pt}{2.5ex}$\frac{2}{1-e^{x}}$\tabularnewline
\hline 
octagon form factor  & \rule{0pt}{2.5ex}$\frac{\cosh y+\cosh\xi}{\cosh y+\cosh\sqrt{x^{2}+\xi^{2}}}$\tabularnewline
\hline 
localization results  & \rule{0pt}{2.5ex}$-\frac{1}{\sinh^{2}\frac{x}{2}}$\tabularnewline
\hline 
\end{tabular}
\par\end{centering}
\caption{The determinant (\ref{eq:detdef}) corresponds to various observables
in four dimensional supersymmetric gauge theories \citep{Beccaria:2022ypy}.
The octagon form factor is related to the four-point function of infinitely
heavy half BPS operators and the parameters $y,\xi$ are the parametrizations
of the conformal cross-ratios. }

\label{Table:symbols} 
\end{table}

The simplest of them is the expectation value of the single trace
half BPS operator ${\rm Tr(Z^{\ell-1})}$ in the presence of a circular
Wilson loop in the maximally supersymmetric gauge theory with $\chi_{w}(x)=-\frac{4\pi^{2}}{x^{2}}$.
The corresponding determinant is simply ~\citep{Beccaria:2023kbl}
\begin{equation}
{\cal D}_{w}(g)=\Gamma(\ell)(2\pi g)^{1-\ell}I_{\ell-1}(4\pi g)
\end{equation}
Its strong coupling expansion has the trans-series with just two terms\footnote{Observe that the logarithm of the determiant would contain infinitely
many trans-series terms having arbitrary integer power of $e^{-8\pi g}$.} 
\begin{align}
{\cal D}_{w}(g)= & Ae^{4\pi g}g^{\frac{1}{2}-\ell}\times\\
\times & \left[\sum_{n\geq0}w_{n}g^{-n}-ie^{i\ell\pi}e^{-8\pi g}\sum_{n\geq0}(-1)^{n}w_{n}g^{-n}\right]\nonumber 
\end{align}
and the non-perturbative part is simply related to the perturbative
one with the relation $w_{n}\to(-1)^{n}w_{n}$, which is also equivalent
to changing the sign of the coupling. Our aim is to present the analogous
trans-series structure for the generic case.

In general, we parametrize the symbol in a Wiener--Hopf type manner
\begin{equation}
1-\chi(x)=bx^{2\beta}\Phi(x)\Phi(-x)\ ;\ \Phi(x)=\prod_{n\geq1}\frac{1-\frac{ix}{2\pi x_{n}}}{1-\frac{ix}{2\pi y_{n}}}
\end{equation}
where $x=-2\pi ix_{n}$ are the zeros, $x=-2\pi iy_{n}$ the poles
of $1-\chi(x)$. They are ordered as $0<x_{i}\leq x_{i+1}$ and $0<y_{i}\leq y_{i+1}$.
The Wilson loop has only one such factor with $x_{1}=1,y_{1}=\infty$
and $\beta=-1$. Normally, we allow arbitrary number of them, some
of them can even coincide and their number can be infinite. We typically
assume the generic situation when they are all different and not even
multiple of each other, and obtain any specific case by the corresponding
smooth limit.

\section{Trans-series expansion}

Based on the previous results \citep{Bajnok2024,Bajnok20251}, concrete
examples and many consistency checks we conjecture that the strong
coupling expansion (valid for $g>0$) has the trans-series structure
\begin{align}
{\cal D}(g)=A & e^{4\pi g\sum_{i}(x_{i}-y_{i})}g^{\frac{\beta^{2}+2\ell\beta}{2}}\times\\
 & \quad\sum_{\{\delta_{i}=0,1\}}e^{-8\pi g\sum_{i}\delta_{i}x_{i}}S^{\{\delta_{i}\}}{\cal D}^{\{\delta_{i}\}}(g)\nonumber 
\end{align}
where ${\cal D}^{\{\delta_{i}\}}(g)$ are asymptotic series in $g^{-1}$,
$e^{-8\pi gx_{i}}$ are non-perturbative scales and $S^{\{\delta_{i}\}}$
are their corresponding Stokes constants. $A$ is a convenient normalization,
which can be calculated from $\chi$ \citep{Beccaria:2022ypy}. The
generic framework to calculate the trans-series for the logarithm
of the determinant already provides a similar structure. The simplification
is that the sum involves only at most once each non-perturbative scale,
so no higher power appears (similarly to the expectation value of
the Wilson loop). In particular, if we have only a finite number of
zeros, then the trans-series contains only a finite number of non-perturbative
corrections.

The logarithm of the perturbative part $\ln D^{\{0,0,\dots\}}(g)$
was already calculated in \citep{Belitsky:2020qrm,Belitsky:2020qir}.
After exponentialization we parametrize it as 
\begin{equation}
{\cal D}^{\{\}}(g)\equiv{\cal D}^{\{0,0,\dots\}}(g)=\sum_{k=0}^{\infty}{\cal D}_{k}^{\{\}}g^{-k}
\end{equation}
by introducing the convention that in the upper index we list the
nonzero $\delta$s only. The coefficients can be calculated from a
system of integro-differential equation with the first few coefficients
being \citep{Belitsky:2020qrm,Belitsky:2020qir}: 
\begin{align}
{\cal D}_{0}^{\{\}} & =1\quad;\quad{\cal D}_{1}^{\{\}}=\frac{1-4\ell_{\beta}^{2}}{16}I_{1}\\
{\cal D}_{2}^{\{\}} & =\frac{16\ell_{\beta}^{4}-40\ell_{\beta}^{2}+9}{512}I_{1}^{2}\quad;\quad\ell_{\beta}=\ell+\beta\nonumber \\
{\cal D}_{3}^{\{\}} & =-\frac{\left(16\ell_{\beta}^{4}-40\ell_{\beta}^{2}+9\right)\left((4\ell_{\beta}^{2}-17)I_{1}^{3}+8I_{2}\right)}{24576}\nonumber 
\end{align}
in terms of the moments: 
\begin{equation}
I_{n}=(-1)^{n-1}\sum_{k\geq1}\left[\frac{1}{(2\pi x_{k})^{2n-1}}-\frac{1}{(2\pi y_{k})^{2n-1}}\right]\label{eq:moments}
\end{equation}
We choose $S^{\{\}}=1$, which fixes $A$ uniquely. Assigning the
degree $n$ to $I_{n}$ we can show that only terms with degree up
to the given order appear. Higher order expressions, ${\cal D}_{k>3}^{\{\}}$,
can be obtained from \citep{Belitsky:2020qrm,Belitsky:2020qir} by
exponentiation.

The leading non-perturbative correction is related to the nearest
zero $x_{1}$, has magnitude $e^{-8\pi gx_{1}}$, and can be parametrized
as 
\begin{equation}
{\cal D}^{\{1\}}(g)\equiv{\cal D}^{\{1,0,\dots\}}(g)=\sum_{k=0}^{\infty}{\cal D}_{k}^{\{1\}}g^{-k}
\end{equation}
Using the methods developed in \citep{Bajnok2024,Bajnok20251} the
coefficients can be determined to be 
\begin{align}
{\cal D}_{0}^{\{1\}} & =1\quad;\quad{\cal D}_{1}^{\{1\}}=\frac{\left(1-4\ell_{\beta}^{2}\right)(\pi x_{1}I_{1}-1)}{16\pi x_{1}},\\
{\cal D}_{2}^{\{1\}} & =\frac{\left(16\ell_{\beta}^{4}-40\ell_{\beta}^{2}+9\right)(\pi x_{1}I_{1}-1)^{2}}{512\pi^{2}x_{1}^{2}}\nonumber \\
{\cal D}_{3}^{\{1\}} & =-\frac{\left(16\ell_{\beta}^{4}-40\ell_{\beta}^{2}+9\right)}{24576\pi^{3}x_{1}^{3}}\biggl(4\ell_{\beta}^{2}(\pi x_{1}I_{1}-1)^{3}+\nonumber \\
 & 8\pi^{3}x_{1}^{3}I_{2}-17\pi x_{1}I_{1}(\pi x_{1}I_{1}(\pi x_{1}I_{1}-3)+3)+19\biggr)\nonumber 
\end{align}
and the corresponding Stokes constant is \citep{Beccaria:2022ypy,Bajnok2024,Bajnok20251}
\begin{align}
S^{\{1\}} & =\frac{e^{i\pi\ell_{\beta}}}{4\pi x_{1}}\frac{\Phi(2i\pi x_{1})}{\Phi'(-2i\pi x_{1})}\\
 & =-ie^{i\pi\ell_{\beta}}\prod_{j>1}\frac{(x_{1}+x_{j})}{(x_{1}-x_{j})}\prod_{j}\frac{(x_{1}-y_{j})}{(x_{1}+y_{j})}\nonumber 
\end{align}
We have calculated many more terms, ${\cal D}_{k}^{\{1\}}$, for $k=4,\dots$
and observed that the perturbative part can be mapped to the leading
non-perturbative correction by the replacements $I_{n}\to I_{n}+\frac{2(-1)^{n}}{(2\pi x_{1})^{2n-1}}$.

This is a very interesting result and has far-reaching consequences.
First, it implies that the two asymptotic series ${\cal D}^{\{\}}$
and ${\cal D}^{\{1\}}$ have the \emph{same perturbative expansion}
in terms of their own moments $I_{n}^{\{\}}\equiv I_{n}$ and $I_{n}^{\{1\}}=I_{n}^{\{\}}+\frac{2(-1)^{n}}{(2\pi x_{1})^{2n-1}}$.
From the definition of the moments it is clear, that if $I_{n}^{\{\}}$
corresponds to $\{x_{1},x_{2},\dots\vert y_{1},y_{2},\dots\}$ then
$I_{n}^{\{1\}}$ corresponds to $\{-x_{1},x_{2},\dots\vert y_{1},y_{2},\dots\}\equiv\{x_{2},\dots\vert x_{1},y_{1},y_{2},\dots\}$.
This nicely matches with the magnitude of the correction $e^{-8\pi gx_{1}}$,
which alltogether with the $e^{4\pi gx_{1}}$ prefactor flips also
the sign of $x_{1}$ in the exponent. Formally we are turning $x_{1}$
to a new $y$. We can then think of ${\cal D}^{\{1\}}$ as the perturbative
part corresponding to the symbol $\Phi^{\{1\}}$ labeled by $\{x_{2},\dots\vert x_{1},y_{1},y_{2},\dots\}$.

In the case of the Wilson loop, when we have only one $x_{1}$, the
transformation $x_{1}\to-x_{1}$ changes the sign of every moment,
which leads to the observed $w_{n}\to(-1)^{n}w_{n}$ relation.

\subsection{Resurgence relations between ${\cal D}^{\{\}}$ and ${\cal D}^{\{1\}}$}

\begin{figure}
\begin{centering}
\includegraphics[width=7cm]{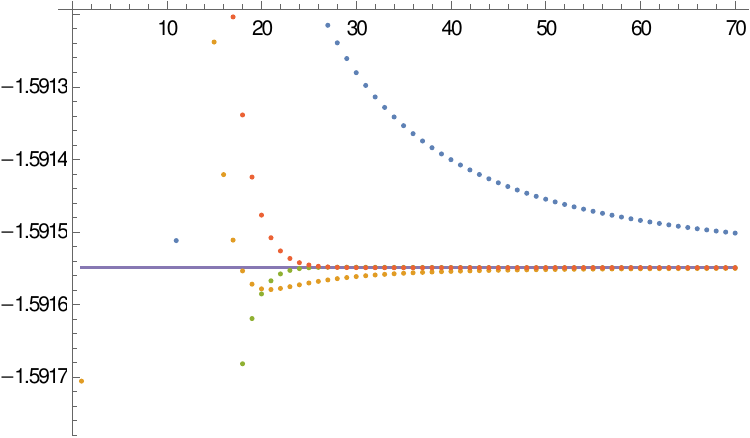}
\par\end{centering}
\centering{}\caption{Asymptotic behaviour of ${\cal D}_{k}^{\{\}}$ for $x_{1}=1,x_{2}=3/2$.
We present the asymptotic series (blue) and its first 3 (yellow, green,
red) Richardson transforms against the asymptotical value $-5/\pi$,
which includes the Stokes constant $S^{\{1\}}$. We zoomed to the
first 70 coefficients for visibility, but we performed the test with
200 terms and also at higher orders. \label{fig:asym12} }
\end{figure}

The perturbative series ${\cal D}^{\{\}}(g)$ and the non-perturbative
correction ${\cal D}^{\{1\}}(g)$ are related by resurgence relations,
see \citep{Marino:2012zq,Dorigoni:2014hea,Aniceto:2018bis} for modern
references. This correspondence originates from ambiguity cancellations
and are manifested in asymptotic relations: the large order behaviour
of the peturbative coefficients ${\cal D}_{k}^{\{\}}$ are related
to the non-perturbative correction ${\cal D}_{k}^{\{1\}}$ as 
\begin{equation}
{\cal D}_{k}^{\{\}}=\sum_{j=0}\frac{S^{\{1\}}\Gamma(k-j+1)}{i\pi(8\pi x_{1})^{k-j}}{\cal D}_{j}^{\{1\}}+\dots\label{eq:asymrel}
\end{equation}
We have checked these relations in many cases. We demonstrate one
of our findings in Figure \ref{fig:asym12} for $x_{1}=1,x_{2}=3/2$. 

Such an asymptotic behaviour determines the nature of the branch cut
at $x_{1}$ of the Borel transform of ${\cal D}^{\{\}}$: 
\begin{equation}
{\cal B}[{\cal D}^{\{\}}](s)=\sum_{k=0}{\cal D}_{k}^{\{\}}\frac{(8\pi s)^{k}}{\Gamma(k+1)}
\end{equation}
We observed that ${\cal B}[{\cal D}^{\{\}}](s)$ does not have any
cut at higher multiple of $x_{1}$. This is in contrast to its logarithm,
where we have cuts and corresponding non-perturbative corrections
at $nx_{1}$ for $n=1,2,\dots$.

\begin{figure}
\begin{centering}
\includegraphics[width=7.5cm]{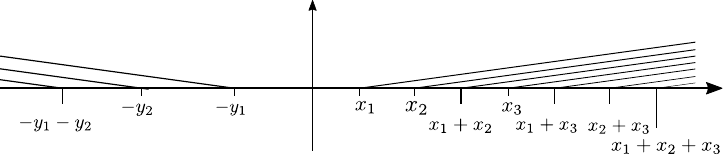} 
\par\end{centering}
\caption{Cut structure on the Borel plane of ${\cal B}[{\cal D}^{\{\}}](s)$
for a symbol with $\{x_{1},x_{2},x_{3}\vert y_{1},y_{2}\}$. We indicated
the branch points of the cuts, while the other branch point is at
infinity. }

\label{fig:Borel0}
\end{figure}

Our generic analysis for the logarithm of the determinant showed that
for every $y_{i}$ a logarithmic cut starts at $-y_{i}$ on the negative
real line, while for any $x_{i}$ a logarithmic cut starts at $x_{i}$
on the positive real line. We investigated numerically many cases
with increasing number of $x_{i}$ s and $y_{i}$s for the determinant
themself and observed that higher cuts appeared only for locations
$\sum_{i}\delta_{i}x_{i}$ with $\delta_{i}=0,1$. This is illustrated
in figure \ref{fig:Borel0} for $\{x_{1},x_{2},x_{3},y_{1},y_{2}\}$.

\begin{figure}
\begin{centering}
\includegraphics[width=7.5cm]{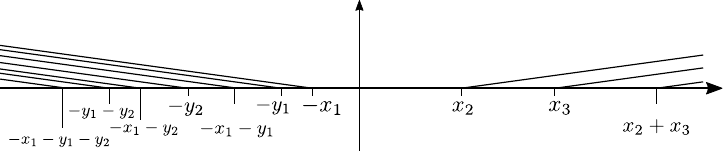} 
\par\end{centering}
\caption{Cut structure on the Borel plane of ${\cal B}[{\cal D}^{\{1\}}](s)$
for a symbol with $\{x_{1},x_{2},x_{3}\vert y_{1},y_{2}\}$. This
is the same cut structure we would have for $\{x_{2},x_{3}\vert x_{1},y_{1},y_{2}\}$.}

\label{fig:Borel1} 
\end{figure}

Since ${\cal D}^{\{1\}}$ corresponds to $\Phi^{\{1\}}$ they are
labeled by $\{x_{2},x_{3},y_{1},y_{2},y_{3}=x_{1}\}$. (Or the $y$s
need to be rearranged according to their magnitudes). The analytic
structure of its Borel transform ${\cal B}[{\cal D}^{\{1\}}](s)$
is represented in figure \ref{fig:Borel1}.

The trans-series so far was formal as the various expansion coefficients
${\cal D}_{k}^{\{\delta_{i}\}}$ are only asymptotical. In order to
give a meaning to the series we use lateral Borel resummation. The
inverse of the Borel transform involves an integration on the real
positive line, where we avoid the branch cuts by integrating slightly
above or below
\begin{equation}
S^{\pm}[{\cal D}^{\{\}}](g)=8\pi g\int_{0}^{\infty e^{\pm i\epsilon}}dse^{-8\pi gs}{\cal B}[{\cal D}^{\{\}}](s)
\end{equation}
 The two lateral Borel resummations are related by the Stokes automorphism
$S^{-}[{\cal D}]=S^{+}[\mathfrak{S}({\cal D})]$. Its logarithm $\ln\mathfrak{S}=\sum_{j}e^{-8\pi gx_{j}}\Delta_{x_{j}}$
contains the alien derivatives, which are related to the jump of the
function at the logarithmic cuts \citep{Marino:2012zq,Dorigoni:2014hea,Aniceto:2018bis}.
We found for the cut at $x_{1}$ that 
\begin{equation}
\Delta_{1}{\cal D}^{\{\}}=-2S^{\{1\}}{\cal D}^{\{1\}}\ ;\quad\Delta_{1}^{2}{\cal D}^{\{\}}=0
\end{equation}
where we used the abbreviation $\Delta_{x_{j}}\equiv\Delta_{j}$ for
$j=1$.

\subsection{Generic relations}

For generic $\{x_{j}\}$ the non-perturbative corrections of order
$e^{-8\pi gx_{j}}$ satisfy the same equations as for $x_{1}$, merely
$x_{1}$ is replaced with $x_{j}$. Thus we can conclude that ${\cal D}^{\{j\}}\equiv{\cal D}^{\{0,\dots,0,1,0,\dots,\}}$
has the same form as ${\cal D}^{\{\}}$ but with $I_{n}^{\{j\}}=I_{n}^{\{\}}+\frac{2(-1)^{n}}{(2\pi x_{j})^{2n-1}}$.
The corresponding Stokes constant is 
\begin{align}
S^{\{j\}} & =\frac{e^{i\pi\ell_{\beta}}}{4\pi x_{j}}\frac{\Phi(2i\pi x_{j})}{\Phi'(-2i\pi x_{j})}\\
 & =-ie^{i\pi\ell_{\beta}}\prod_{m\neq j}\frac{(x_{j}+x_{m})}{(x_{j}-x_{m})}\prod_{m}\frac{(x_{j}-y_{m})}{(x_{j}+y_{m})}
\end{align}
and the alien derivatives are 
\begin{equation}
\Delta_{j}{\cal D}^{\{\}}=-2S^{\{j\}}{\cal D}^{\{j\}}\ ;\quad\Delta_{j}^{2}{\cal D}^{\{\}}=0
\end{equation}
We can think of ${\cal D}^{\{j\}}$ as the perturbative part corresponding
to $\Phi^{\{j\}}$. Since $\Phi^{\{j\}}$ has no zero at $x_{j}$
(rather a pole there) we do not expect any singularity on the Borel
plane at $x_{j}$ neither a non-perturbative correction at order $e^{-8\pi x_{j}g}$
or any higher multiple of them. Indeed, this is what we found by analysing
the asymptotic behaviour of ${\cal D}_{k}^{\{j\}}$ as well as the
analytic structure of ${\cal B}[{\cal D}^{\{j\}}](s)$ in many cases.

However, the non-perturbative corrections ${\cal D}^{\{j\}}$ will
still have Borel singularities related to $x_{k}$ and its asymptotic
behaviour indicates corrections of the form $e^{-8\pi x_{k}g}$. Since
our argumentations work for generic $\{x_{1},\dots,x_{j},\dots\vert y_{1},y_{2},\dots\}$
it will also work for $\{x_{1},\dots,\hat{x}_{j},\dots\vert x_{j,}y_{1},y_{2},\dots\}$,
where with the hat we indicated that $x_{j}$ is absent. Thus, following
from the resurgence relations (\ref{eq:asymrel}), we can say that
${\cal D}^{\{j,k\}}\equiv{\cal D}^{\{0,\dots0,1,0,\dots,0,1,0,\dots\}}$
must have the same expansion as the perturbative part but with the
replacements $I_{n}\to I_{n}^{\{j,k\}}$, where 
\begin{align}
I_{n}^{\{j,k\}} & =I_{n}^{\{j\}}+\frac{2(-1)^{n}}{(2\pi x_{k})^{2n-1}}=I_{n}^{\{k\}}+\frac{2(-1)^{n}}{(2\pi x_{j})^{2n-1}}
\end{align}
The corresponding Stokes constants are 
\begin{align}
S^{\{j,k\}} & =S^{\{j\}}S^{\{k\}}\frac{(x_{j}-x_{k})^{2}}{(x_{j}+x_{k})^{2}}
\end{align}
where the extra factor is due to the fact that we need to take $\frac{e^{i\pi\ell_{\beta}}}{4\pi x_{k}}\frac{\Phi^{\{j\}}(2i\pi x_{k})}{\Phi^{\{j\}\prime}(-2i\pi x_{k})}$
rather than the same expression with $\Phi$, which defines $S^{\{k\}}$.
Observe that the result does not depend on which order we take the
non-perturbative corrections. Stating in a different manner 
\begin{equation}
\Delta_{j}\Delta_{k}{\cal D}^{\{\}}=\Delta_{k}\Delta_{j}{\cal D}^{\{\}}
\end{equation}

Repeating the same procedure further we can see that ${\cal D}^{\{j_{1},\dots,j_{n}\}}$
has exactly the same expansion as ${\cal D}^{\{\}}$, but we have
to replace $I_{n}$ with $I_{n}^{\{j_{1},\dots,j_{n}\}}$, which is
obtained from $I_{n}$ by flipping the signs of $x_{j_{k}}$s 
\begin{equation}
I_{n}^{\{j_{1},\dots,j_{l}\}}=I_{n}+\sum_{k=1}^{l}\frac{2(-1)^{n}}{(2\pi x_{j_{k}})^{2n-1}}
\end{equation}
Similarly, the Stokes constants are 
\begin{equation}
S^{\{j_{1},\dots,j_{l}\}}=\prod_{m}S^{\{j_{m}\}}\prod_{m<n}\frac{(x_{j_{m}}-x_{j_{n}})^{2}}{(x_{j_{m}}+x_{j_{n}})^{2}}
\end{equation}
and the alien derivatives read as 
\begin{equation}
\Delta_{k}{\cal D}^{\{j_{1},\dots,j_{l}\}}=-2\frac{S^{\{k,j_{1},\dots,j_{l}\}}}{S^{\{j_{1},\dots,j_{l}\}}}{\cal D}^{\{k,j_{1},\dots,j_{l}\}}
\end{equation}
from which it follows that different alien derivatives commute $\Delta_{j}\Delta_{k}{\cal D}^{\{j_{1},\dots,j_{l}\}}=\Delta_{k}\Delta_{j}{\cal D}^{\{j_{1},\dots,j_{l}\}}$.
It is also true that $\Delta_{k}^{2}{\cal D}^{\{j_{1},\dots,j_{l}\}}=0$.

By slightly changing the normalization of the expansion function 
\begin{equation}
S^{\{\delta_{j}\}}{\cal D}^{\{\delta_{j}\}}(g)=\prod_{j}(S^{\{j\}})^{\delta_{j}}\bar{{\cal D}}^{\{\delta_{j}\}}
\end{equation}
the determinant can be formulated as an infinite parameter trans-series
\begin{align}
{\cal D}(g,\{\sigma\})=A & e^{4\pi g\sum_{i}(x_{i}-y_{i})}g^{\frac{\beta^{2}+2\ell\beta}{2}}\times\\
 & \quad\sum_{\{\delta_{i}=0,1\}}e^{-8\pi g\sum_{i}\delta_{i}x_{i}}(\prod_{i}\sigma_{i}^{\delta_{i}})\bar{{\cal D}}^{\{\delta_{i}\}}(g)\nonumber 
\end{align}
Since the alien derivaties are proportional to derivatives wrt. the
trans-series parameters $\Delta_{j}{\cal D}(g,\{\sigma\})=-2\partial_{\sigma_{j}}{\cal D}(g,\{\sigma\})$
the Stokes automorphism, which connects the two lateral resummations,
shifts the trans-series parameters as $\sigma_{j}\to\sigma_{j}+2S^{\{j\}}$,
and the median resummation provides the physical answer: $\sigma_{j}=S^{\{j\}}$.

\subsection{Example}

In the case of $\{x_{1},x_{2},x_{3}\vert y_{1},\dots\}$ the non-perturbative
corrections can be represented by a 3 dimensional cube. The vertex
$\{\}$ represents the perturbative part, ${\cal D}^{\{\}}$, while
each other vertex represents a non-perturbative correction $,{\cal D}^{\{1\}},\dots$.
The oriented edges show non-trivial resurgence relations between them.
The alien derivatives move in the corresponding directions and add
their index to the set.

In the general case of the symbol $\{x_{1},\dots,x_{l}\vert y_{1},\dots\}$
the non-perturbative corrections and resurgence relations can be visualized
on an $l$-dimensional cube. 
\begin{figure}
\begin{centering}
\includegraphics[width=4cm]{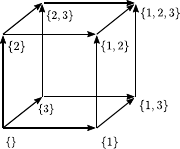} 
\par\end{centering}
\caption{Non-perturbative corrections for $\{x_{1},x_{2},x_{3}\vert y_{1},\dots\}$
and their resurgence relations. }
\end{figure}

\section{Conclusions}

In this Letter we derived the complete trans-series for semi-infinite
determinants associated with the Bessel kernel, generalizations of
the Tracy--Widom distributions that appear in many contexts, including
observables of ${\cal N}=4$ super Yang--Mills theory. We found that
the non-perturbative corrections exhibit a special structure: each
non-perturbative scale enters only linearly, and the corrections can
be labeled by a set of integers $\{j_{k}\}$ specifying which scales
$e^{-8\pi x_{j_{k}}g}$ are present. All perturbative series multiplying
a given correction share the same universal form, expressed in terms
of transformed moments $I_{n}^{\{j_{1},\dots,j_{k}\}}$. These moments
are obtained from the symbol by a simple rule that converts the selected
zeros $x_{j_{k}}$ into poles. Our result is based on the relation
connecting the perturbative sector to the leading non-perturbative
correction, which we have tested to high order and against resurgence
consistency conditions. 

To our knowledge, the strikingly simple resurgence structure we found
has not appeared previously in the literature. It would be interesting
to understand whether it reflects a deeper non-perturbative mechanism
in supersymmetric gauge theories and to extend the analysis to other
observables, such as the tilted cusp \citep{Basso:2020xts,Bajnok20252}.
A selection of symbols directly corresponding to gauge theory observables
is listed in (\ref{Table:symbols}), where we plan to explore in detail
the physical meaning of the associated non-perturbative corrections.

\section{Acknowledgements}

The research was supported by the Doctoral Excellence Fellowship Programme
funded by the National Research Development and Innovation Fund of
the Ministry of Culture and Innovation and the Budapest University
of Technology and Economics, under a grant agreement with the National
Research, Development and Innovation Office (NKFIH). DlP acknowledges
support by the Alexander von Humboldt Foundation.

\bibliographystyle{elsarticle-num}
\bibliography{references}

\end{document}